\documentclass[a4paper,11pt]{article}
\usepackage{graphicx,amssymb,amstext,amsmath,mathabx,mathtools,esvect}

\usepackage{floatflt}
\usepackage{subcaption,ragged2e}
\usepackage[dvipsnames]{xcolor}

\definecolor{cbl}{rgb}{0,0,1}                

\newcommand{\bc}{\begin{center}}
\newcommand{\ec}{\end{center}}
\def\ba#1{\begin{array}{#1}\displaystyle}
\newcommand{\ea}{\end{array}}

\newcommand{\beq}{\begin{equation}}
\newcommand{\eeq}{\end{equation}}
\newcommand{\beqa}{\begin{eqnarray}}
\newcommand{\eeqa}{\end{eqnarray}}

\newcommand{\bi}{\begin{itemize}}
\newcommand{\ei}{\end{itemize}}

\newcommand{\bra}{\langle}
\newcommand{\ket}{\rangle}

\newcommand{\TT}{{\cal T}}

\usepackage{cite}

\definecolor{purple_nice}{rgb}{0.4,0.2,0.7}
\definecolor{fuel_blue}{RGB}{42,162,185}
\definecolor{YInMn_blue}{RGB}{46, 80, 144}
\definecolor{ultramarine}{RGB}{63, 0, 255}
\definecolor{KLEIN_blue}{rgb}{0, 0.18, 0.65}
\usepackage[linktocpage=true]{hyperref}
\hypersetup{
    colorlinks=true,
    linkcolor=YInMn_blue,
    citecolor=ultramarine,
    filecolor=fuel_blue,
    urlcolor=KLEIN_blue,
}

\begin{document}

\begin{titlepage}
\title{A Toy Model for Topological Entanglement Features \\in 1+1D Integrable Quantum Field Theory}
\author{Emmalise J. H. Aalbers{\color{green} $^\diamondsuit$} and Olalla A. Castro-Alvaredo{\color{red} {$^\heartsuit$}}}
\date{\small 
{\color{green} {$^\diamondsuit$}}  Institute for Theoretical Physics Utrecht University,\\
Princetonplein 5, 3584 CC Utrecht, The Netherlands\\
\medskip{\color{red} {$^\heartsuit$}}  Department of Mathematics, City St George's, University of London,\\ 10 Northampton Square EC1V 0HB, UK\\
\medskip
}
\maketitle
\begin{abstract}
In this paper we investigate an entanglement measure, the R\'enyi entropy, in a 1+1D integrable quantum field theory known as the Federbush model. This is a deformation of the theory of two massive Dirac fermions by means of a bilinear term in the $U(1)$ currents that couples the two fermion species. This deformation gives rise to $S$-matrix elements which are coupling-dependent phases, distinct from $- 1$. These non-trivial phases can be seen as encoding anyon-like statistics. From this viewpoint, the Federbush model is a toy model for topological features of entanglement in one space dimension. In this paper we show that, a standard computation based on form factors of branch point twist fields leads to the conclusion that these topological features play no role when computing most known measures of entanglement at equilibrium in the ground state. This conclusion applies also to the post-quench dynamics at first order in perturbation theory, after a small quench of the topological parameter.

\end{abstract}

\bigskip
\bigskip
\noindent {\bfseries Keywords:} Integrable Quantum Field Theory, Federbush Model, Topology, Entanglement, Form Factors

\vfill

\noindent 
{\Large {\color{green} {$^\diamondsuit$}}} e.j.h.aalbers@students.uu.nl\\
{\Large{\color{red} {$^\heartsuit$}}} o.castro-alvaredo@city.ac.uk \\

\hfill \today
\end{titlepage}
\section{Introduction}
In recent years the properties of entanglement measures, particularly their growth properties with respect to subsystem size and time, have emerged as prime signatures of distinct types of quantum systems: integrable, many-body localized, chaotic etc. While well-known area and volume laws \cite{arealaw,volumelaw} characterize the growth of entanglement as a function of subsystem size at equilibrium \cite{Latorre1,Jin,Calabrese:2004eu,entropy}, linear \cite{quench1,quench2,higherD,Rottoli:2025axj,IsingQuench,Maulik:2025blg,Rakovszky:2019qjb}, logarithmic \cite{MBL} and other kinds of growth \cite{powerlaw,Agarwal_2015,Vosk_2014,nonhermitian}, characterize the time-evolution of entanglement following a quantum quench. 

Topology is another important aspect of quantum systems and of quantum states which has been studied in the context of entanglement. In particular, there is a notion of topological entanglement entropy \cite{TEE,Levin_2006,Wen_2017}, that is associated with a next-to-leading order constant correction to the  entanglement entropy (as a function of subsystem size), similar to the boundary entropy of 1+1D conformal field theories \cite{Affleck:1991tk,Cardy:1989ir}. It is also known that, topological features of quantum states, such as the topological order, are particularly salient when studying the so-called entanglement hamiltonian \cite{ES,Gu_2009}, whose eigenvalues characterize the entanglement spectrum \cite{Serbyn:2016xun}. While these measures reveal important information about the structure of quantum states, there is also a complementary notion of topological quantum computation \cite{Lahtinen:2017ynm} where the unusual exchange statistics of excitations known as anyons can be harnessed to achieve fault-tolerant quantum computation processing \cite{Pachos_2012}. 

In this paper, we study a standard entanglement measure, the R\'enyi entropy, both at and away from equilibrium in a 1+1D massive integrable quantum field theory (IQFT). The chosen theory, known as Federbush model 
\cite{Feder1,Feder2}, has an extremely simple scattering matrix, which can be seen as encoding anyon statistics in one space dimension. It is an ideal candidate for exploring markers of topology in 1+1D gapped systems.  We will now set the scene by introducing the main computation techniques and the entanglement measure of interest. 

\medskip

Let $|\Psi\ket$ be a pure state of an IQFT and let us define a bipartition of space into two complementary regions $A$ and $\bar{A}$ so that the Hilbert space of the theory $\mathcal{H}$ also decomposes into a direct product $\mathcal{H}_A \otimes \mathcal{H}_{\bar{A}}$. Then the reduced density matrix associated to subsystem $A$ is obtained by tracing out the degrees of freedom of subsystem $\bar{A}$ in
\beq
\rho_A=\mathrm{Tr}_{\bar{A}}(|\Psi\ket \bra \Psi|)\,,
\eeq
and the von Neumann and $n$th R\'enyi entropy of a subsystem $A$ are defined as
\beq
S=-\mathrm{Tr}_A(\rho_A \log \rho_A)\quad \mathrm{and} \quad S_n=\frac{\log(\mathrm{Tr}_A \rho_A^n)}{1-n}\,.
\label{SS}
\eeq
where $\mathrm{Tr}_A \rho_A^n:={\mathcal{Z}}_n/{\mathcal{Z}}_1^n$ can be interpreted as the normalised partition function of a theory constructed from $n$ non-interacting copies or replicas of the original model. As is well known, $S=\lim_{n\rightarrow 1} S_n$. 
\medskip

It has been known for some time that the partition function $\mathrm{Tr}_A \rho_A^n:={\mathcal{Z}}_n/{\mathcal{Z}}_1^n$ of the replica theory can be identified with a correlation function of local fields in QFT \cite{Calabrese:2004eu} known as branch point twist fields (BPTFs) \cite{entropy}. In particular, if the theory is massive, the correlators can be obtained by expanding the correlation functions into sums over form factor contributions \cite{entropy}. This technique has been employed very successfully over the past 20 years both for equilibrium \cite{entropy,ourboundary,FBoson,CLevi} and out-of-equilibrium \cite{Oscillations,IsingQuench,SGDavid} situations and also much earlier for other fields in IQFT (see ie \cite{Fring_1993,Z,Yurov:1990,Fring:1992pj,Luky1,Castro-Alvaredo_2001, CastroAlvaredo:2000nk, CastroAlvaredo:2000em,CastroAlvaredo:2000ag,CastroAlvaredo:2000nr,takacs}) since its inception in the 70s \cite{KW,smirnov1992book}. Denoting by $\TT$ and $\tilde{\TT}$ the branch point twist field and its hermitian conjugate, we have that, at equilibrium, for a subsystem of length $\ell$ within an infinite system, $\mathrm{Tr}_A \rho_A^n \sim {}_n\bra 0| \TT(0)\tilde{\TT}(\ell)|0\ket_n$, where $|0\ket_n$ is the ground state of the replica theory and we consider a same-time correlator. Consequently, as per (\ref{SS}) the R\'enyi entropy is a function of the two-point function of BPTFs. Similarly, if we consider instead an out-of-equilibrium situation, and a bipartition into two semi-infinite regions, then $\mathrm{Tr}_A \rho_A^n \sim {}_n\bra 0| \TT(t)|0\ket_n$, where we now consider the field at the origin of space. 
\medskip 

One of the key advantages of integrability is that both the two-point and one-point functions above can, a priori at least, be computed exactly by expanding them in terms of matrix elements of the corresponding fields. These matrix elements are called form factors and can be defined as
\beq 
F_{(a_1,j_1)\ldots (a_k,j_k)}^n(\theta_1,\ldots,\theta_k):={}_n\bra 0| \TT(0)|\theta_1,\ldots,\theta_k|0\ket_n\,,
\label{FF0}
\eeq 
where $\theta_i$ are the usual rapidity variables in terms of which energy and momenta of a particle of mass $m$ are parametrized $e(\theta):=m \cosh\theta$, $p(\theta)=m\sinh\theta$. $|0\ket_n$ is the ground state and $|\theta_1,\ldots,\theta_k|0\ket_n$ is a $k$-particle excited state in the replica theory. The indices $(a_i,j_i)$ label each particle species $a_i \in [1,N]$ and the copy $j_i \in [1,n]$ of the theory where it resides. The two-point correlation function then admits a standard large-distance expansion in terms of these form factors
\beqa 
&& {}_n\bra 0| \TT(0)\tilde{\TT}(\ell)|0\ket_n \\
&& 
=\sum_{k=0}^\infty \frac{1}{k!(2\pi)^{k} } \sum_{a_1,\ldots, a_{k}=1}^{N}\sum_{j_1,\ldots, j_{k}=1}^n \prod_{i=0}^k \int_{-\infty}^\infty d\theta_i  \, \left| F_{(a_1,j_1)\ldots (a_k,j_k)}^n(\theta_1,\ldots,\theta_k)\right|^2 e^{-\ell \sum_{i=1}^{
k}m_i \cosh\theta_i},\nonumber
\label{correlation}
\eeqa 

This paper is organized as follows: in Section \ref{federbush} we introduce the Federbush model. We present its lagrangian and $S$-matrix and argue that it represents the simplest integrable model which admits a topological interpretation. In Section \ref{BPTF} we compute the form factors of the BPTF and show that, due to the form of the scattering matrix and the form factor equations, the form factors do not depend on the topological parameter. In Section \ref{Quench} we focus on the case of a global quench of the topological parameter and use first-order quench perturbation theory to compute the time evolution of the one-point function of the BPTF. We show that at this order, the one-point function remains unchanged after the quench. We present our conclusions and outlook in Section \ref{conclusion}. 
\section{The Federbush Model}
\label{federbush}
The Federbush model is a relativistic IQFT, first proposed and studied in the 60s \cite{Feder1, Feder2} and further investigated in \cite{current, Castro-Alvaredo_2001,pseudoscalar, cadamuro}.  The theory consists of two massive stable particles and their conjugates. Let us label these particles as $\{1,2,\bar{1},\bar{2}\}$. Then the two-body scattering matrices can be written as:
\beq 
S_{1 2}(\theta)=S_{\bar{1}\bar{2}}(\theta)=S_{\bar{2}{1}}(\theta)=S_{{2}\bar{1}}(\theta)=-e^{-2\pi i \lambda}\,,
\label{fed1}
\eeq 
\beq  S_{21}(\theta)=S_{\bar{2}\bar{1}}(\theta)=S_{\bar{1}{2}}(\theta)=S_{{1}\bar{2}}(\theta)=-e^{2\pi i \lambda}\,,
\label{fed2}
\eeq 
and
\beq 
S_{11}(\theta)=S_{1\bar{1}}(\theta)=S_{\bar{1}1}(\theta)=S_{\bar{1}\bar{1}}(\theta)=S_{22}(\theta)=S_{2\bar{2}}(\theta)=S_{\bar{2}2}(\theta)=S_{\bar{2}\bar{2}}(\theta)=-1\,,
\label{Sff}
\eeq 
where $\lambda \in (0,2)$, $\lambda \neq 1$ and $\theta$ is the rapidity variable, even though, in this case, the scattering amplitudes are rapidity independent. The coupling $\lambda$ also features in the lagrangian of the theory which takes the form:
\beq 
\mathcal{L}= \sum_{a=1,2} \bar{\Psi}_a (i\gamma^\mu \partial_\mu - m_a)\Psi_a - 2\pi \lambda \epsilon_{\mu\nu} J_1^\mu J_2^\nu\,,
\label{feder}
\eeq 
where $\Psi_a$ is a field associated with a massive particle of mass $m_a$ and $J_a^\mu=\bar{\Psi}_a\gamma^\mu \Psi_a$ are the associated vector currents. As usual $\epsilon_{\mu \nu}$ is the Levi-Civita antisymmetric tensor and $\gamma^\mu$ are matrices satisfying the Clifford algebra. An important property of the lagrangian is that it is $U(1)$ invariant $\Psi_a \mapsto \eta_a \Psi_a$, where $\eta_a$ is a $U(1)$ phase. It is clear that, for $\lambda=0$ we recover a theory of two free Dirac fermions, in accordance with the $S$-matrices (\ref{Sff}). 

The Federbush model has two features that are relatively rare in IQFT. First, it breaks parity, namely, in general $S_{ab}(\theta)\neq S_{ba}(\theta)$. Second, $S_{ab}(0)\neq \pm 1$ for generic $a, b$. Both of these properties are quite unique and only found in a family of models with more complicated interactions, that is the homogeneous sine-Gordon (HSG) models \cite{hsg,FernandezPousa:1997iu,ntft,CastroAlvaredo:2000ag,CastroAlvaredo:2000nr}. Indeed, it has been shown that the Federbush model can be obtained as a particular limit of a HSG model \cite{Castro-Alvaredo_2001}. 

In 1+1D quantum field theory, interaction and particle statistics are hard to separate since particles cannot be exchanged without undergoing a scattering process. In that sense the scattering matrix can itself be seen as the statistical factor. In the Federbush model this feature is particularly apparent because the scattering matrix is rapidity-independent, it is a phase which can be directly interpreted statistically. Hence, the $S$-matrix can be seen as encoding anyon-like statistics in 1D.  The phases $-e^{\pm 2\pi i \lambda}$ encode both the fermionic origin of the theory (the $-1$) and the interaction (the scattering phase $e^{-2\pi i \lambda}$). Several examples of exactly solvable non-relativistic models with anyonic statistics exist, notably \cite{Schulz_1998,Kundu:1998mp}.
 
It is interesting to ask if and how entanglement measures will depend on the coupling $\lambda$. In this paper we will answer this question both at equilibrium and following a small global quantum quench of the parameter $\lambda$ \cite{quench1,quench2}. Our main conclusion is that, both the standard equilibrium measures and the entanglement entropy following a global quench of the parameter $\lambda$ are $\lambda$-independent and therefore carry no information about the topological phases. These conclusions are somewhat surprising, at least from the viewpoint of the technique used which depends directly on the $S$-matrix. However, they have some precedent in the existing literature, as we discuss in the conclusion.

\section{Branch Point Twist Field Form Factors}
\label{BPTF} 
We have defined form factors in the Introduction as matrix elements of local fields between the ground state and an excited state. In IQFTs a systematic program to compute them has existed for a long time \cite{KW,smirnov1992book} and this program was later generalized in \cite{entropy} to deal with branch point twist fields. As discussed in the same paper, BPTFs are symmetry fields associated to cyclic permutation symmetry. We usually denote this field by $\TT$ and its hermitian conjugate by $\tilde{\TT}$. This symmetry arises when one considers a replica version of the model under investigation. Let $n$ be the replica number, to simplify notation slightly, we will define the $k$-particle form factors of BPTFs as 
\beq 
F_{a_1\ldots a_k}^n(\theta_1,\ldots,\theta_k):={}_n\bra 0| \TT(0)|\theta_1,\ldots,\theta_k|0\ket_n\,,
\label{FF}
\eeq 
when all particles live in the same copy of the theory (so we can drop the copy numbers $j_i$ from the original definition (\ref{FF0})). 
It is known that form factors involving particles in different copies, are related to the above via rapidity shifts. However, as we shall see, we will not need those in this paper. 

The form factors (\ref{FF}) can be computed by solving a set of equations that specify their monodromy and analyticity properties. We will not list all these equations here (they can be found in many papers, including \cite{entropy}). The fact that we are dealing with an ``almost free" theory will simplify the solution procedure considerably. We will only present the few equations that are critical for our current problem and refer the reader for instance to \cite{CLevi} for a more detailed description of the solution procedure. 

We start by observing that due to $U(1)$ symmetry of the Lagrangian,  the only non-vanishing form factors are those involving equal numbers of any particle $a=1,2$ and its conjugate $\bar{a}$. The computation of form factors typically starts with the computation of minimal form factors (MFFs), which are entire solutions to the two-particle form factor equations in the physical strip, in our case ${\rm Im} (\theta) \in [0, n \pi]$. Due to Lorentz invariance, one can argue that they are functions of a single variable $F_{ab}^n(\theta_1,\theta_2)\propto f^n_{ab}(\theta_{12})$. The proportionality means that there will be in general another function of $\theta_{12}:=\theta_1-\theta_2$ involved, which will encode the pole structure of the full form factor and will, in particular, ensure that the full form factor is zero if $b\neq \bar{a}$. The MFFs satisfy the equations,
\beq 
f^n_{ab}(\theta)=S_{ab}(\theta)f^n_{ba}(-\theta)=f^n_{ba}(-\theta+2\pi i n)\,.
\label{ffeq}
\eeq 
where $a,b$ are particle indices for particles living in the same copy of an $n$-replica theory and we take these particles to be arbitrary, not necessarily conjugated to each other, since we will need all MFFs to construct higher-particles form factors. As we see from the $S$-matrices (\ref{Sff}), if $a=\bar{b}$ or $a=b$ then the corresponding MFF is the same as for the free fermion theory, which is known \cite{entropy} 
\beq 
f^n_{a\bar{a}}(\theta)=f^n_{aa}(\theta)=-i \sinh\frac{\theta}{2n}\,\qquad \mathrm{for}\qquad a=1,2,\bar{1},\bar{2}\,.
\eeq 
On the other hand, if we now consider two-particle combinations whose scattering matrices involve the non-trivial phase we have the equations
\beqa 
f^n_{12}(\theta)&=&-e^{-2\pi i \lambda} f^n_{21}(-\theta)=f^n_{21}(-\theta+2\pi i n)\,,\nonumber\\
 f^n_{21}(\theta)&=&-e^{2\pi i \lambda} f^n_{12}(-\theta)=f^n_{12}(-\theta+2\pi i n)\,.
\label{MFF12}
\eeqa 
These are essentially the same equations that are satisfied by $U(1)$ fields in the Dirac theory, thus not entirely new \cite{Bernard:1994re}. It is easy to see that 
\beq 
f_{12}^n(\theta)=f^n_{\bar{1}\bar{2}}(\theta)=f^n_{\bar{2}{1}}(\theta)=f^n_{{2}\bar{1}}(\theta)= -e^{-2\pi i\lambda} \, e^{\frac{\theta}{n} \left(\lambda+\frac{1}{2}\right)}\,,
\eeq 
and
\beq  
f_{21}^n(\theta)=f^n_{\bar{2}\bar{1}}(\theta)=f^n_{\bar{1}{2}}(\theta)=f^n_{{1}\bar{2}}(\theta)=e^{-\frac{\theta}{n} \left(\lambda+\frac{1}{2}\right)}\,,
\eeq 
are solutions of the MFF equation (\ref{MFF12}). The normalization of these functions is not too important since the full form factor is normalized by further requirements we will see below. 

Form factors involving a particle and antiparticle of opposite momenta will have a kinematic pole, a property that is encapsulated by the kinematic residue equation. For BPTFs this equation is very simple 
\beq 
{\rm Res}_{\theta_0=\theta} F^n_{\bar{a}a a_1\ldots a_k}(\theta_0+i\pi,\theta,\theta_1,\ldots, \theta_k)= i F^n_{a_1\ldots a_k}(\theta_{1},\ldots,\theta_k)\,.
\label{KRE}
\eeq 
Due to the replica structure, it is also known that for BPTFs there must be a corresponding kinematic pole with opposite residue at $\theta_0-\theta=i\pi(2n-1)$. The full non-vanishing two-particle form factors are
\beq 
F^n_{\bar{a}{a}}(\theta)= -i \frac{ \tau_n {\cos\frac{\pi}{2n}}\sinh\frac{\theta}{2n}}{n\sinh\left(\frac{i\pi +\theta}{2n}\right) \sinh\left(\frac{i\pi-\theta}{2n}\right)}\,\qquad \mathrm{for}\qquad a=1,2\,,
\label{FF2}
\eeq 
which are identical to those already known for free fermions \cite{entropy,nexttonext}, now extended to complex free fermions. Here $\tau_n:={}_n \bra 0| \TT(0) |0 \ket_n$ is the vacuum expectation value of the field. Since these are the only non-vanishing two-particle form factors, this means that, at the two-particle order at least, the BPTF carries no information at all about the topological nature of the $S$-matrix and any computations of entanglement measures in the two-particle approximation (that is, involving only two-particle form factors, or up to $k=2$ in the expansion \eqref{correlation}) will give (four times) the Majorana fermion/Ising result.

{ One may ask whether there is further freedom in the form factor equations that would allow us to obtain solutions distinct from (\ref{FF2}), possibly $\lambda$-dependent. Typically the same form factor equations may have multiple solutions corresponding to different operators. In the present case, it is possible to argue that the solution (\ref{FF2}) is the correct one. There are two reasons for this: first, it is known that the ultraviolet conformal dimension of the BPTF is proportional to the central charge of the UV conformal field theory \cite{Calabrese:2004eu,entropy}. For the Federbush model this is $c=2$ since it is a perturbation of two massless Dirac fermions. It is possible to obtain the conformal dimension of our BPTF by employing the $\Delta$-sum rule \cite{DSC} and it should be exactly four times the result for the Ising field theory, which was obtained in \cite{entropy}. Furthermore, the $\Delta$-sum rule computation requires knowing the form factors of the trace of the stress-energy tensor, which were obtained in \cite{Castro-Alvaredo_2001} and shown to be identical to those of the Ising model (just like (\ref{FF2})). In particular only two-particle form factors are non-vanishing so that higher particle form factors do not contribute to the $\Delta$ sum rule. It follows then that applying the $\Delta$-sum rule with the form factors (\ref{FF2}) will deliver the correct conformal dimension. Any modifications of the form factor above would have to be additive terms that have no kinematic poles and which give zero contribution to the $\Delta$-sum rule. For this to happen, such terms would need to be even functions of $\theta$ which is incompatible with the first form factor equation (\ref{ffeq}), therefore there can not be any such additional terms at two-particle level. }

The two-particle result does not mean however, that the $\lambda$-dependent MFFs computed above are not useful. They will feature in computations of higher particle form factors, where the topological nature of the theory might reveal itself. The simplest situation where this could happen is in the four-particle form factors $F^n_{\bar{a}{a}\bar{b}b}(\theta_1,\theta_2,\theta_3,\theta_4)$ with $a,b=1,2$. We will now turn our attention to these.
\subsection{Computing the 4-Particle Form Factors}
As is known from the usual construction, higher-particle form factors also have a minimal part and a pole structure, and if we focus just on the non-vanishing 4-particle form factors $F^n_{\bar{a}a\bar{b}b}(\theta_1,\theta_2,\theta_3,\theta_4)$  we can write
\beq 
F^n_{\bar{a}a\bar{b}b}(\theta_1,\theta_2,\theta_3,\theta_4)=H_{\bar{a}a\bar{b}b}^n Q_{\bar{a}a\bar{b}b}^n (\theta_1,\theta_2,\theta_3,\theta_4) P_{\bar{a}a\bar{b}b}^n(\theta_1,\theta_2,\theta_3,\theta_4)\,, 
\eeq 
where 
\beqa  
P_{\bar{a}a\bar{b}b}^n(\theta_1,\theta_2,\theta_3,\theta_4)& =&  \frac{f^n_{\bar{a}a}(\theta_{12})f^n_{\bar{b}b}(\theta_{34})}{\sinh\left(\frac{i\pi +\theta_{12}}{2n}\right) \sinh\left(\frac{i\pi-\theta_{12}}{2n}\right) \sinh\left(\frac{i\pi +\theta_{34}} {2n}\right) \sinh\left(\frac{i\pi-\theta_{34}}{2n}\right)}\label{ansatz} \\
&& \times \frac{f^n_{\bar{a}\bar{b}}(\theta_{13})f^n_{\bar{a}b}(\theta_{14})f^n_{{a}\bar{b}}(\theta_{23})f^n_{ab}(\theta_{24})}{\left(\sinh\left(\frac{i\pi +\theta_{14}}{2n}\right) \sinh\left(\frac{i\pi-\theta_{14}}{2n}\right) \sinh\left(\frac{i\pi +\theta_{23}}{2n}\right) \sinh\left(\frac{i\pi-\theta_{23}}{2n}\right)\right)^{\delta_{ab}}}\,, \nonumber
\eeqa  
are chosen so as to automatically solve Watson's equations (thanks to the MFFs in the numerator) and to incorporate all the required poles (thanks to the $\sinh$-factors in the denominator). 

Here $H_{\bar{a}a\bar{b}b}^n $ are constants and $Q_{\bar{a}a\bar{b}b}^n $ are symmetric and $2\pi i n$-periodic functions of the rapidities. They are determined by the corresponding form factor residue equation (\ref{KRE}),
\beq 
{\rm Res}_{\theta_0=\theta} F^n_{\bar{a}a\bar{b}b}(\theta_0+i\pi,\theta,\theta_1,\theta_2)= i F^n_{\bar{b}b}(\theta_{12})\,.
\label{15}
\eeq 
In this case, because of the simple structure of the two-particle form factors it is actually quite easy to solve for the unknowns $H_{\bar{a}a\bar{b}b}^n $ and $Q_{\bar{a}a\bar{b}b}^n $. In the case when $a=b$ we can use known results for free fermions and argue that the solution has a Pfaffian structure, as exploited in \cite{ourboundary}. Furthermore, in that case, there will again be no dependence on the parameter $\lambda$. Therefore, the only new original case is $a\neq b$.  Consider for example $a=1$, $b=2$. Then,
\beqa  
P_{\bar{1}1\bar{2}2}^n(\theta_1,\theta_2,\theta_3,\theta_4)=  \frac{f^n_{\bar{1}1}(\theta_{12})f^n_{\bar{2}2}(\theta_{34})f^n_{\bar{1}\bar{2}}(\theta_{13})f^n_{\bar{1}2}(\theta_{14})f^n_{{1}\bar{2}}(\theta_{23})f^n_{12}(\theta_{24})}{\sinh\left(\frac{i\pi +\theta_{12}}{2n}\right) \sinh\left(\frac{i\pi-\theta_{12}}{2n}\right) \sinh\left(\frac{i\pi +\theta_{34}}{2n}\right) \sinh\left(\frac{i\pi-\theta_{34}}{2n}\right)}\,,
\eeqa

First we compute
\beq
{\rm Res}_{\theta_0=\theta} P^n_{\bar{1}1\bar{2}2}(\theta_0+i\pi,\theta,\theta_1,\theta_2)= \frac{i n^2 e^{-4 \pi i \lambda} }{\cos^2\frac{\pi}{2 n}} F^n_{\bar{2}{2}}(\theta)\,. 
\label{res}
\eeq 
Comparing with (\ref{15}) we see that we can simply take the polynomial $Q^n_{\bar{1}1\bar{2}2}(\theta_1,\theta_2,\theta_3,\theta_4)=1$ and 
\beq 
H^n_{\bar{1}1\bar{2}2}= -i \frac{e^{4\pi i \lambda} \cos^2\frac{\pi}{2n}}{n^2}\,.
\eeq 
Interestingly, we have that the $\lambda$ dependence in the residue (\ref{res}) is due to the property 
\beq 
f^n_{\bar{1}\bar{2}}(\theta_{13})f^n_{\bar{1}2}(\theta_{14})f^n_{{1}\bar{2}}(\theta_{23})f^n_{12}(\theta_{24})=e^{-4\pi i \lambda}\,,
\label{simp}
\eeq 
so, any dependence on the topological parameter is canceled out due to the normalization required by the kinematic residue equation. In fact, we find that, both for $a=b$ and for $a\neq b$ there is a very simple factorized solution into free fermion contributions
\beq
F^n_{\bar{a}a\bar{b}b}(\theta_1,
\theta_2,\theta_3,\theta_4)= F^n_{\bar{a}{a}}(\theta_{12}) F^n_{\bar{b}{b}}(\theta_{34})\,. 
\label{fac}
\eeq
This means that a computation of the entanglement entropy of an interval will not reveal information about the topological phase, even at order four in the form factor expansion. The conclusion, is even stronger in fact since, repeated use of the kinematic residue equation and an ansatz of the type \eqref{ansatz}, leads to the conclusion that the factorization property (\ref{fac}) occurs for any particle numbers. { In particular, the form factor $F^n_{\bar{1}1\ldots\bar{1}1\bar{1}\bar{2}2\ldots \bar{2}2}(\theta_1, \ldots,\theta_{4k})$ will be proportional to the product of minimal form factors
\beqa 
&& \prod_{i=1}^{2k} f_{\bar{1}1}(\theta_{i\,i+1})\prod_{i=2k+1}^{4k} f_{\bar{2}2}(\theta_{i\,i+1})\nonumber\\
&& \prod_{i=1}^{k} \prod_{j=1}^{k} f_{\bar{1}\bar{2}}(\theta_{2i-1\,2k+2j-1}) f_{{1}\bar{2}}(\theta_{2i\,2k+2j-1}) f_{{1}{2}}(\theta_{2i\,2k+2j}) f_{\bar{1}{2}}(\theta_{2i-1\,2k+2j})\,,
\label{25}
\eeqa 
where we see that the product in the second line is exactly of the same type as 
(\ref{simp}) and therefore, equation (\ref{25}) equals:
\beq 
e^{-8k \pi i \lambda} \prod_{i=1}^{2k} f_{\bar{1}1}(\theta_{i\,i+1})\prod_{i=2k+1}^{4k} f_{\bar{2}2}(\theta_{i\,i+1})
\eeq 
and once more the $\lambda$ dependence is canceled out by the normalization of the form factors imposed by the kinematic residue equation, giving the structure}
\beq
F^n_{\bar{1}1\ldots\bar{1}1\bar{1}\bar{2}2\ldots \bar{2}2}(\theta_1, \ldots,\theta_{4k})= F^n_{\bar{1}{1}\ldots \bar{1}{1}}(\theta_{1}\ldots \theta_{2k}) F^n_{\bar{2}{2}\ldots\bar{2}{2} }(\theta_{2k+1},\ldots,\theta_{4k})\,.
\label{higher}
\eeq
Since the form factors in the factorization involve only particles that interact as free fermions they are $\lambda$-independent. These form factors are known in the literature \cite{entropy,ourboundary}. In summary, a computation of the R\'enyi entropy or von Neumann entropy of an interval of length $\ell$ by the formula \eqref{correlation}, will give exactly the same result as for two Dirac fermions, with no trace of the topological phase. In fact, since most known measures can be written as correlations of BPTFs, this conclusion applies to other measures of entanglement too, such as the entropies of disconnected regions \cite{disco1,disco2,disco4} and the logarithmic negativity \cite{negativity1,negativity2,negativity3,negativity4,ourneg}. 

{ Similar to the case of two-particle form factors, we can ask if there are additional terms with no kinematic poles that could be added to the formulae above and which would introduce a $\lambda$ dependence. While this can not be excluded at the level of the form factor equations, there are arguments as to why this is not possible. First, the solution (\ref{higher}) follows directly from the two-particle form factor (\ref{FF2}) by using the kinematic residue equation, so whichever other solutions are possible, they will have to include this contribution plus potentially additional terms that introduce a $\lambda$ dependence. Second, because this $\lambda$-dependence does not come from the seed two-particle form factor, it would have to be introduced in an ad hoc, nonphysical manner. Finally, from the arguments presented after equation (\ref{FF2}), if such solutions existed they would correspond to fields of exactly the same UV conformal dimension as the field with form factors (\ref{higher}). This is incompatible with uniqueness of the BPTF which is defined as the unique symmetry field associated to cyclic permutation symmetries of lowest conformal dimension.}

\section{Quench Perturbation Theory}
\label{Quench}
Quench perturbation theory can be employed to investigate the time dependence of the one-point function of the BPTF. Our aim in this section is to ascertain whether the time-evolution of the one-point function displays a dependence in the coupling $\lambda$.   
In \cite{Pertur} a perturbation theory  for  the study of integrable models subject to a small integrability-breaking perturbation was developed.  The case considered there was translation invariant in time, which is no longer the case in a non-equilibrium protocol, such as a quench. In \cite{Delfino:2014qfa} an approach to tackle the quench problem was proposed in which the state in the Heisenberg picture after the quench is expanded perturbatively in the quench parameter over the pre-quench quasi-particle basis. The approach requires the pre-quench theory to be integrable but allows for considering integrability-breaking protocols. 

Let us review the main results of \cite{Delfino:2014qfa}, adapting them to our particular case. Consider the theory of two Dirac fermions with ground state $|0\ket$ and action $\mathcal{A}_0$.  At time $t=0$ the system is quenched and is thereafter described by the perturbed action
\beq
\label{action}
\mathcal{A}=\mathcal{A}_0-2\pi \lambda \int_0^\infty dt \int_{-\infty}^\infty dx \, \texttt{J}^2(x,t)\, \quad \mathrm{with}\quad \texttt{J}^2:=\epsilon_{\mu \nu} J_1^\mu J_2^\nu\,.
\eeq
 In the interaction picture, with  respect to the Hamiltonian of the pre-quench theory, the state of the system at infinite time after the quench is the time-ordered exponential
\begin{equation}
\label{state_p}
|\psi_0\ket= \lim_{t\rightarrow\infty} T \left[\exp\left(-2\pi i \lambda \int_0^t ds \int_{-\infty}^\infty dx \, \texttt{J}^2(x,s) \right)\right]|0\ket\,. 
\end{equation}
The state $|\psi_0\rangle$ in~\eqref{state_p}  can then be expanded perturbatively in $\lambda$ over the basis of the out-states of the pre-quench theory. $k$-particle states of this type  are denoted by  $|\theta_1,\ldots, \theta_k|0 \ket$,  with $\theta_1<\theta_2 < \cdots <\theta_k$, being the rapidities. It can then be shown that   
\beqa
|\psi_0\rangle=|0\ket + 2\pi \lambda \sum_{k=1}^\infty  \frac{2\pi}{k!} \int_{-\infty}^\infty  \prod_{i=1}^k \frac{d \theta_{i}}{2\pi}~\frac{ \delta(\sum_i \hat{p}_{a_i}(\theta_i)) [F_{a_1\ldots a_k}^{\texttt{J}^2}(\theta_1,\ldots,\theta_k)]^*}{\sum_{i} \hat{e}_{a_i}(\theta_i)} |\theta_1\ldots \theta_k|0\ket+O(\lambda^2)\,, 
\label{psi0}
\eeqa
represents the state in the pre-quench basis in the Heisenberg picture at all times after the quench, up to first order in $\lambda$. $\hat{e}_a(\theta)=\hat{m}_a \cosh \theta$ and $\hat{p}_a(\theta)=\hat{m}_a \sinh \theta$ are the pre-quench energy and momenta of the particles,
 and
\beq
F_{a_1\ldots a_k}^{\texttt{J}^2}(\theta_1,\ldots,\theta_k):=\bra 0| \texttt{J}^2(\bold{0})|\theta_1,\ldots,\theta_k|0\ket\,,
\label{J2FF}
\eeq
is a $k$-particle form factor of the local field $\texttt{J}^2$, calculated in the pre-quench quasi-particle basis. Once we know the form factors, the state above can then be employed  to compute perturbative corrections to the one-point function of any local field after the quench, including the field $\TT$. We will return to this computation, once we have discussed how to compute  (\ref{J2FF}).

\subsection{The Form Factors of the Field $\texttt{J}^2$}
Our aim in this subsection is to compute the non-vanishing form factors of  the field $\texttt{J}^2$ introduced in \eqref{action}. Let us write the field more explicitly as
\beq 
\texttt{J}^2=\epsilon_{\mu \nu} J_1^\mu J_2^\nu =J_1^0 J_2^1-J_1^1 J_2^0\,.
\label{J}
\eeq 
If we see this field as a perturbation around the free fermion point, we have that, from the definition $J_a^\mu=\bar{\psi}_a\gamma^\mu \psi_a$, each current is a bilinear on Dirac fermions
\beq 
\psi_a(\texttt{x})= \int \frac{d\theta}{\sqrt{4\pi}} (\texttt{a}_a(\theta) u_a(\theta) e^{-i p_a(\theta) \cdot \texttt{x}} + \texttt{a}^\dagger_{\bar{a}}(\theta) v_a(\theta) e^{i p_{{a}}(\theta)\cdot \texttt{x}})\,,
\eeq 
where $\bar{\psi}_a= \psi^\dagger_a \gamma^0$,
\beq 
\psi^\dagger_a(\texttt{x})= \int \frac{d\theta}{\sqrt{4\pi}} (\texttt{a}^\dagger_a(\theta) u^\dagger_a(\theta) e^{i p_a(\theta) \cdot \texttt{x}} + \texttt{a}_{\bar{a}}(\theta) v^\dagger_a(\theta) e^{-i p_{{a}}(\theta)\cdot \texttt{x}})\,,
\eeq 
and $\texttt{x}=(x^0,x^1)$, $p_a=(m_a \cosh\theta, m_a \sinh\theta)$. The creation-annihilation operators satisfy the fermion algebra 
\beq 
\{\texttt{a}_a(\theta),\texttt{a}_b(\beta)\}=0, \qquad \{\texttt{a}_a(\theta),\texttt{a}^\dagger_b(\beta)\}=2\pi \delta_{ab} \delta(\theta-\beta)\,,
\label{anti}
\eeq 
and we have the usual definitions of the Clifford matrices and spinors:
\beq
\gamma_0= \left(\begin{array}{cc}
0& 1\\
1& 0
\end{array}\right)\,, \quad \gamma_1=\left(\begin{array}{cc}
0& 1\\
-1& 0
\end{array}\right)\,, \quad \gamma^0\gamma^0=\left(\begin{array}{cc}
1& 0\\
0& 1
\end{array}\right)\,,\quad \gamma^0 \gamma^1= \left(\begin{array}{cc}
-1& 0\\
0& 1
\end{array}\right)\,,
\eeq 
\beqa 
&& u_a(\theta)=\sqrt{\frac{m_a}{2}} \left(\begin{array}{c}
e^{-\frac{\theta}{2}}\\
e^{\frac{\theta}{2}}
\end{array}\right)\,, \quad v_a(\theta)=i\sqrt{\frac{m_a}{2}} \left(\begin{array}{c}
e^{-\frac{\theta}{2}}\\
-e^{\frac{\theta}{2}}
\end{array}\right)\,,\nonumber\\
&& u^\dagger_a(\theta)=\sqrt{\frac{m_a}{2}} \left(\begin{array}{cc}
e^{-\frac{\theta}{2}}& 
e^{\frac{\theta}{2}}
\end{array}\right)\,,\quad v^\dagger_a(\theta)=-i\sqrt{\frac{m_a}{2}} \left(\begin{array}{cc}
e^{-\frac{\theta}{2}}&
-e^{\frac{\theta}{2}}
\end{array}\right)\,.
\eeqa
Since $\texttt{J}^2$ is quartic in fermions, its only non-vanishing, independent form factor is
\beq 
\bra 0| \texttt{J}^2(0)|\texttt{a}_{\bar{1}}^\dagger(\theta_1) \texttt{a}_1^\dagger(\theta_2)\texttt{a}_{\bar{2}}^\dagger(\theta_3) \texttt{a}_2^\dagger(\theta_4)|0\ket\,,
\label{FFJ}
\eeq 
which, taking into account the anticommutation relations (\ref{anti}) and the fact that annihilation operators annihilate the vacuum, means that only the annihilation operator part of the fermionic fields will contribute to computing (\ref{FFJ}).
At $\texttt{x}=0$ we have simply
\beqa 
J_a^0= {\psi}^\dagger_a\gamma^0 \gamma^0 \psi_a =\int \frac{d\theta}{\sqrt{4\pi}} (\texttt{a}^\dagger_a(\theta) u^\dagger_a(\theta)  + \texttt{a}_{\bar{a}}(\theta) v^\dagger_a(\theta) ) \int \frac{d\beta}{\sqrt{4\pi} } (\texttt{a}_a(\beta) u_a(\beta) + \texttt{a}^\dagger_{\bar{a}}(\beta) v_a(\beta) )\,. 
\eeqa 
and
\beqa 
J_a^1= {\psi}^\dagger_a \gamma^0 \gamma^1 \psi_a = \int \frac{d\theta}{\sqrt{4\pi} } (\texttt{a}^\dagger_a(\theta) u^\dagger_a(\theta) + \texttt{a}_{\bar{a}}(\theta) v^\dagger_a(\theta)) 
\int \frac{d\beta}{\sqrt{4\pi} } (i \texttt{a}_a(\beta) v_a(\beta) - i \texttt{a}^\dagger_{\bar{a}}(\beta) u_a(\beta))\,. 
\eeqa 
Putting everything together, after some simple algebra, we get 
\beqa 
&& \bra 0| \texttt{J}^2(0)|\texttt{a}_{\bar{1}}^\dagger(\theta_1) \texttt{a}_1^\dagger(\theta_2)\texttt{a}_{\bar{2}}^\dagger(\theta_3) \texttt{a}_2^\dagger(\theta_4)|0\ket = \int \frac{d\beta_1}{\sqrt{4\pi} } \int \frac{d\beta_2}{\sqrt{4\pi}} \int \frac{d \beta_3}{\sqrt{4\pi}} \int \frac{d \beta_4}{\sqrt{4\pi}}  \nonumber\\ 
&& \bra 0| \texttt{a}_{\bar{1}}(\beta_1) \texttt{a}_1 (\beta_2) \texttt{a}_{\bar{2}}(\beta_3) \texttt{a}_2 (\beta_4)| g(\beta_1,\beta_2,\beta_3,\beta_4)  |\texttt{a}_{\bar{1}}^\dagger(\theta_1) \texttt{a}_1^\dagger(\theta_2)\texttt{a}_{\bar{2}}^\dagger(\theta_3) \texttt{a}_2^\dagger(\theta_4)|0\ket\,,
\eeqa 
with 
\beq 
g(\beta_1, \beta_2,\beta_3,\beta_4):=i \left[ v_1^\dagger(\beta_1)u_1(\beta_2) v_2^\dagger(\beta_3)v_2(\beta_4)-v_1^\dagger(\beta_1)v_1(\beta_2) v_2^\dagger(\beta_3)u_2(\beta_4)\right]\,. 
\eeq 
Using the relations (\ref{anti}) we have
\beqa 
&& \bra 0|\texttt{a}_{\bar{1}}(\beta_1) \texttt{a}_1 (\beta_2) \texttt{a}_{\bar{2}}(\beta_3) \texttt{a}_2 (\beta_4)|\texttt{a}_{\bar{1}}^\dagger(\theta_1) \texttt{a}_1^\dagger(\theta_2)\texttt{a}_{\bar{2}}^\dagger(\theta_3) \texttt{a}_2^\dagger(\theta_4)|0\ket\nonumber\\
&& = (2\pi)^4 \delta(\beta_1-\theta_1) \delta(\beta_2-\theta_2)   \delta(\beta_3-\theta_3)  \delta(\beta_4-\theta_4)\,,
\eeqa 
and 
\beqa 
 \bra 0| \texttt{J}^2(0)|\texttt{a}_{\bar{1}}^\dagger(\theta_1) \texttt{a}_1^\dagger(\theta_2)\texttt{a}_{\bar{2}}^\dagger(\theta_3) \texttt{a}_2^\dagger(\theta_4)|0\ket &=& \pi^2 g(\theta_1,\theta_2,\theta_3,\theta_4)=-\pi^2 m^2 \sinh\frac{\hat{\theta}_{12}-\hat{\theta}_{34}}{2}\,,
\eeqa 
with $\hat{\theta}_{12}:=\theta_1+\theta_2$ and $\hat{\theta}_{34}:=\theta_3+\theta_4$.
\subsection{Post-Quench Time  Evolution}
We now have all necessary ingredients to continue our perturbative computation. In our case, the state \eqref{psi0} is simplified by the fact that only one independent form factor is non-vanishing. This allows us to get rid of the sum in $k$ and write 
\beqa
|\psi_0\rangle = |0\ket + \lambda  \frac{\pi^2}{6} \int_{-\infty}^\infty  \prod_{i=1}^4 \frac{d \theta_{i}}{2\pi}~\frac{ \delta(\sum_{i=1}^4 \hat{p}(\theta_i)) [F_{\bar{1} 1 \bar{2} 2}^{\texttt{J}^2}(\theta_1,\theta_2,\theta_3,\theta_4)]^*}{\sum_{i=1}^4 \hat{e}(\theta_i)} |\theta_1\ldots \theta_4|0\ket +O(\lambda^2)\,, 
\label{psi0simple}
\eeqa
where we assume that the masses of the two fermions are identical $m_1=m_2=m$, so that we can remove particle indices in the energy and momenta. We also use $\hat{m}$ for the pre-quench mass and $m$ for the post-quench mass (in this section). 

Then, the change of the one-point function of the BPTF at first order in perturbation theory can be written as
\beqa
\delta \tau_n(t) := {}_n\bra \psi_0|\TT(0,t)|\psi_0 \ket_n -{}_n \bra 0|\TT(0,0)|0\ket_n = \frac{ n \lambda \pi^2}{6} I(t,n) +C_\TT+O(\lambda^2)\,, 
\label{delta}
\eeqa
where 
\beq 
I(t;n):=\int_{-\infty}^\infty \prod_{i=1}^4 \frac{d \theta_{i}}{2\pi}~\frac{ \delta(\sum_{i=1}^4 \hat{p}(\theta_i)) }{\sum_{i=1}^4 \hat{e}(\theta_i)}2\mathrm{Re}\left[[F_{\bar{1}1\bar{2}2}^{\texttt{J}^2}(\theta_1,\theta_2,\theta_3,\theta_4)]^* F^n_{\bar{1}1\bar{2}2}(\theta_1,\theta_2,\theta_3,\theta_4) e^{-i \sum_{i=1}^4 \hat{e}(\theta_i) t}\right] \,, 
\eeq 
and $C_\TT= -\frac{n\lambda \pi^2}{6} I(0,n)$~\cite{Delfino:2016bln}
is a constant which ensures that $\delta \tau_n(0) =0 $ at first order in perturbation theory. Note the factor $n$, which accounts for considering $n$ copies of the theory.

This approach has been employed successfully in a number of works, notably \cite{Oscillations,SGDavid} where undamped oscillations were found which relate to the presence of non-vanishing one-particle form factors and which were confirmed numerically (see e.g. \cite{EK}).
These have been described in further generality within this perturbative approach in \cite{peros}. 
Perturbation theory was also employed for describing the oscillations of entanglement in the Ising model after a mass quench \cite{IsingQuench}, in which case oscillations are damped even at first order in perturbation theory. The approach has also been generalised to dealing with excited states \cite{EStates}.

As noticed above, the present case is simpler because the expansion of the state (\ref{psi0}) at first order in perturbation theory contains only a single form factor contribution. However, it is also more complicated in that this single form factor involves 4 particles, while all the examples mentioned above considered 1- or 2-particle form factors only. Some simplifications are nonetheless possible. For example, changing integration variables to $x:=\theta_{12}, y:=\theta_{34}$ and $u:=\frac{\hat{\theta}_{12}}{2}, v:=\frac{\hat{\theta}_{34}}{2}$  we have that $I(t,n)$ can be written as 
\beqa 
I(t;n) &=&\frac{1}{(2\pi)^4}\int_{-\infty}^\infty dx \int_{-\infty}^\infty dy \int_{-\infty}^\infty du \int_{-\infty}^\infty dv~\frac{ \delta(2\hat{m}(\cosh\frac{x}{2}\sinh u + \cosh\frac{y}{2} \sinh v)) }{2\hat{m}(\cosh\frac{x}{2}\cosh u + \cosh\frac{y}{2} \cosh v)} \nonumber\\
&& 2\mathrm{Re}\left[\mathcal{F}(x,y,u,v) e^{-2\hat{m} i t (\cosh\frac{x}{2}\cosh u + \cosh\frac{y}{2} \cosh v)}\right] \,, 
\eeqa 
with 
\beq 
\mathcal{F}(x,y,u,v):=  \pi^2 \hat{m}^2 F_{\bar{1}1}(x) F_{\bar{2}2}(y) \sinh(u-v)\,. 
\eeq 
A crucial feature of the integral is that, while the sum of energies and the momentum constraint  are both invariant under the transformation  $(u,v) \mapsto (-u,-v)$, the function $\mathcal{F}(x,y,u,v)$ picks up a minus sign. Since the transformation also leaves the Jacobian invariant, this means that the integral must vanish, purely on symmetry grounds, and so:
\beq 
\delta\tau_n(t) =0
\eeq 
Therefore, at least at first order in perturbation theory, the anyonic part of the $S$-matrix leaves no trace in the post-quench dynamics. 

\section{Conclusion and Outlook}
\label{conclusion}
In this paper we considered the Federbush model, a 1+1D integrable quantum field theory which has anyonic exchange relations in one space dimension. Our aim was to investigate how and if the topological properties of the theory had an effect on various entanglement measures, both at and away from equilibrium. 
Our main conclusions are:
\begin{enumerate}
\item Any equilibrium measures in the ground state, which can be expressed in terms of correlators of branch point twist fields, are independent of the topological parameter. They will be the same as for a pair of decoupled massive Dirac fermions. 
\item At first order in perturbation theory, following a global quench of the topological parameter, the entanglement entropy of a semi-infinite system remains unchanged w.r.t. to its equilibrium value in the Dirac theory. 
\end{enumerate}
Given that all our calculations are based on the form factors of BPTFs and that these form factors satisfy equations that involve the full $S$-matrix, these conclusions are surprising. We found that the form factors of BPTFs, unlike those of other fields in the theory \cite{Castro-Alvaredo_2001}, have a factorized structure which is independent of the topological parameter. Hence any entanglement measures built from these form factors will have no dependence either. In the case of quench perturbation theory, the symmetry properties of the form factors of the BPTF and of the chosen perturbing  field, dictate that the first correction in perturbation theory is vanishing. 

These conclusions are  less surprising from the physical viewpoint, and in fact, similar results have been found for other examples, although, as far as we know, not in IQFT. Some arguments are: first, that at least in non-relativistic theories, anyonic models in one space dimension can be mapped to free theories \cite{Girardeau_2006}, therefore their local properties are the same as for those; second, that at least in the massless UV limit both the double massive Dirac theory and the Federbush model flow to the same CFT with $c=2$, therefore, in the conformal limit at least, they both have the same entanglement properties; third, and related to the first point, that the exchange of 1D anyons is not an independent topological operation because in 1D exchange and scattering are the same thing. The intuition would then be that the topological parameter should play a role in nonlocal properties (like the correlation functions of at least some local fields) but not on local ones (like entanglement). An interesting recent example where the post-quench dynamics  is also found to be independent of the topological parameter can be found here \cite{quench3}. 

\medskip

Given our conclusions, it is worth asking whether there are any other interesting settings that we could consider, where the anyonic part of the $S$-matrix can leave an imprint on entanglement measures. Once more, existing literature for non-relativistic theories, suggests that there are \cite{AnyonRing1,AnyonRing2}. One way to achieve this might be to consider the theory in finite volume. Even though the problem of computing the finite-volume form factors of BPTFs is still open, for free theories at least, this can be solved as shown in \cite{excited,excited1,SRexcited}. It is likely that a similar approach can be employed for the Federbush model and that the anyonic phases then enter the computation via the quantization conditions of the momenta in finite volume. This is related to the problem of how to generalize the (thermodynamic) Bethe ansatz to deal with Haldane-Wu statistics \cite{Bytsko:1998iq,Bytsko:2002fy}.
Another situation where we are likely to see a dependence in $\lambda$ is for other global quenches. For instance, we may consider a quench of the Federbush model by a different field of the theory,  a field whose form factors depend explicitly on $\lambda$. Examples of such fields and their form factors can be found in \cite{Castro-Alvaredo_2001}. Finally, we could consider other measures of entanglement which are based on different types of BPTFs, such as symmetry resolved measures. This is natural for this model since there is  $U(1)$ symmetry. It is likely that the form factors of these ``composite" BPTFs will depend explicitly on $\lambda$.  We hope to return to these problems in future work.

\medskip
\noindent {\bf Acknowledgments:} Emmalise J. H.~Aalbers is grateful to the Department of Mathematics of City St George's, University of London, for hosting her during her undergraduate research project and this follow up project. Olalla A. Castro-Alvaredo is grateful to the mathematical research institute MATRIX in Creswick (Australia) where part of this research was done and to Zlatko Papi\'c for a useful email exchange, which included  bringing the papers \cite{AnyonRing1,AnyonRing2,Schulz_1998} to her attention. 

\bibliography{bibliography}
\end{document}